# A French Connection? Recognition and Entente for the Taliban

Authors- Mandeep Singh Rai Misty Wyatt, Sydney Farrar, and Kristina Alabado
Affiliation- The Fletchers School of Law and Diplomacy, Tufts University.
Corresponding authors email address- Mandeep.Rai@tufts.edu,

**Introduction**

This paper will explore a way forward for French-Afghan relations post United Nations (U.N.) occupation. The summer of 2021 proved to be very tumultuous as the Taliban lay in waiting for U.N. forces to withdraw in what many would call a hasty, poorly thought-out egress operation. The Taliban effectively retook all major cities, including the capital, Kabul, and found themselves isolated as U.N. states closed embassies and severed diplomatic relations. Now, Afghanistan finds itself without international aid and support and on the verge of an economic crisis. France now has the opportunity to initiate a leadership role in the establishment and recovery of the new Afghan government.

The French-Afghan relationship began in the early twentieth century but remained tenuous for decades. The U.N. occupation offered an opportunity for change through diplomatic relations and economic investments. In a broad sense, Afghanistan's general opening and modernization during the U.N. occupation does not align with the Taliban that is today in power. However, the group now has an immediate interest in establishing relations to facilitate economic and humanitarian aid. France is uniquely positioned to coordinate and lead any possible humanitarian initiatives.

France, being recently scorned by the United States (U.S.) and Australia, two close allies, in the deal to purchase submarines, may be eager to show the U.S. they are and intend to remain a global power. In addition, leading humanitarian efforts and broader diplomatic conversations could earn France significant clout in the European Union and globally. Thus, in the long term, France would influence Afghanistan's future relationships to France's benefit. This holds the possibility of placing the U.S., once the lead occupier of Afghanistan, in the backseat of Afghanistan's, and potentially the region's future.

There will be two positions suggested in this paper. The first will be a recommendation for an immediate solution addressing the most urgent of issues facing the Afghan people. France will need to ensure humanitarian conditions are clear and honored. If satisfactorily achieved, France will also provide aid under the watchful eye of a third-party observer to ensure aid reaches those in need and under humanitarian guidelines.

The second, long term recommendation is for officially established French-Afghan relations with the continuation of humanitarian efforts. In concert with these efforts, the Taliban's stance on terrorist groups that reside within the state and abroad must be addressed. This in turn will offer legitimacy to the Taliban as Afghanistan's government both internationally and domestically. Additionally, recognition will open the door to international organization support by way economic and security matters. If all of these conditions are met, the standard of life within Afghanistan will increase and both parties will benefit.

**Overview of the Taliban**

The Taliban as an organization was born from an ultra-fundamentalist sect within Sunni Islam, Deobandi. They emerged in 1994 as a militant group in the middle of the Afghan Civil War (1992-1996) led by veterans of the Soviet-Afghan War (1978-1992) from the Southeastern provinces. In 1996, they conquered Kabul, establishing the Islamic Emirates of Afghanistan. The Taliban took control of nearly 75% of the country, only withstanding the Northern Alliance, a small resistance faction holding against Taliban fundamentalist rule. In two years' time, the Taliban had seized control of over 90% of the country. Due to their grotesque human rights violations, only three states recognized the Taliban as a legitimate government: United Arab Emirates, Saudi Arabia, and Pakistan.[1]

Under Taliban rule, Sharia Law was the jurisprudence that carried severe consequences. Women were subjected to a range of restrictions, including wearing full body coverings, requiring a male relative escort in public, and not being allowed to seek an education or work outside of the home. Violations of such laws were punishable by public beating or even death.[2] Similar consequences were afflicted upon Christians, Hindus, and Sikhs. Music, art, television, movies, radio, media were all banned. Cultural genocide was prevalent, and massacres were common to maintain fear-based control of the population.

Because the Taliban had diplomatic ties with only three states, trade was minimal. The Golden Crescent[3] was the source of revenue for the Taliban, primarily for the sale and trafficking of heroin. Heroin, derived from poppies,[4] is a natural resource of Afghanistan. Due to a lack of infrastructure for extraction and exportation, the country was, and remains, unable to economically benefit from vast deposits of copper and iron. Human trafficking and kidnapping for ransom were also sources of revenue for the Taliban.[5] Even with revenue streams from illicit means, the country was in economic ruins. Saudi Arabia was the primary supplier of desperately needed aid.

From the end of the Afghan Civil War, a Sunni extremist group, Al-Qaeda, found refuge under Taliban rule in Afghanistan. Taliban allowed the terrorist network to train and grow unhindered because they found commonality in fundamental Sunni derived views and a shared history from the Soviet-Afghan War. Provided a haven by the Taliban, Al-Qaeda was effectively elevated as a "state-sponsored" terrorist group. Al-Qaeda's large-scale, international attacks initiated the invasion of Coalition Forces into Afghanistan to destroy the network of terrorists and remove the Taliban from power.[6]

---

[1] "Who are the Taliban?," *BBC*, August 18, 2021, https://www.bbc.com/news/world-south-asia-11451718.
[2] William Maley, *Fundamentalism Reborn?: Afghanistan and the Taliban* (London: Hurst, 2001), 145–166
[3] Neamatollah Nojumi, in *The Rise of the Taliban in Afghanistan: Mass Mobilization, Civil War, and the Future of the Region* (Basingstoke: Palgrave, 2009), 178.
[4] Pierre-Arnaud Chouvy, *Opium: Uncovering the Politics of the Poppy* (Cambridge, MA: Harvard University Press, 2010), 52.
[5] Hanif Sufizada Education and Outreach Program Coordinator, "The Taliban Are Megarich – Here's Where They Get the Money They Use to Wage War in Afghanistan," *The Conversation*, August 23, 2021, https://theconversation.com/the-taliban-are-megarich-heres-where-they-get-the-money-they-use-to-wage-war-in-afghanistan-147411.

After the twenty year-long occupation of Coalition Forces, the Taliban swiftly reclaimed the country in an astonishing eleven days. Yet, the group finds themselves in a financial crisis without partners to trade or provide urgently needed aid. Moreover, the Taliban claims they do not want to be an exiled state (once again) and have assured the international community that they will not institute such atrocities as they inflicted historically.

While most countries have severed official diplomatic relations with the Taliban, many unofficial communications remain. The Taliban understands the need to establish relationships; however, they balance this with establishing sovereignty. Of the significant economic and political powers in the world, France, as a leader within the European Union, has a great deal of influence on the international stage. Therefore, a pragmatic compromise or stalemate between France and the Taliban could benefit the citizens of Afghanistan.

**French-Afghan Relations**

French and Afghan relations began in 1922 and have experienced a series of turmoils from the Soviet invasion in 1979 to the country's civil war to the present Taliban "2.0". Nevertheless, despite objections from the U.S., France was the first Western power to announce the reopening of its embassy in Kabul after the Soviet withdrawal.[7] Likewise, when NATO took the lead of the International Security Assistance Force (ISAF) in 2003, France and Germany were the first two countries to expand ISAF's presence in Afghanistan northward.[8]

---

[6] "Afghanistan Wakes after Night of Intense Bombings," *CNN*, accessed October 3, 2021, http://edition.cnn.com/2001/US/10/07/gen.america.under.attack/.

Since autumn 2001, France was directly involved in operations to fight terrorism, secure the country, and train Afghan security forces. Its intervention is based on Security Council resolutions and the North Atlantic Treaty, covering individual and collective self-defense, and cooperation in the fight against international terrorism. As part of Operation Enduring Freedom, France initially sent special forces under American command for specific missions, Ares and Héraclès, in the fight against terrorism in the region. The French forces were mainly based in the Northeast of Kabul and operated from the airbases of Bagram and Kandahar. Since 2002, within the framework of Operational Mentoring and Liaison Teams (OMLT), France assisted in the reconstitution of the Afghan national army through training and joint missions.[9]

France has also been heavily involved in the reconstruction of Afghanistan, starting with the 2002 action program designed to respond to humanitarian emergency and to relaunch traditional cooperation: aid to the Esteqlal and Malalai French-speaking high schools in Kabul, collaboration in the fields of health, agriculture, culture, governance, and the rule of law.[10] In 2004, the decision to include Afghanistan in the Priority Solidarity Zone (PSZ) allowed the country to benefit from financing from the French Development Agency (AFD) and access to the Priority Solidarity Fund (PSF).[11]

---

[7] Elaine Sciolino, "France Plans to Reopen Kabul Embassy," *The New York Times,* January 10, 1990 (accessed October 3, 2021); available from https://www.nytimes.com/1990/01/10/world/france-plans-to-reopen-kabul-embassy.html.
[8] *International Security Assistance Force (ISAF)* (accessed October 3, 2021); available from https://www.understandingwar.org/international-security-assistance-force-isaf.
[9] Zalmaï Haquani, "Les relations entre la France et l'Afghanistan," *Académie de Géopolitique de Paris,* April 7, 2016, http://www.academiedegeopolitiquedeparis.com/les-relations-entre-la-france-et-lafghanistan/.
[10] *French Aid for Reconstruction* (accessed September 21, 2021); available from https://www.amb-afg.fr/en/aide_francaise.

France also coordinated international efforts to support the establishment of the Afghan Parliament, in narrow collaboration with the United Nations Development Programme (UNDP) and its partners, particularly European. The French Ministry of Foreign Affairs allocated EUR 2.5 million and the European Commission EUR 3.5 million for this project, which consisted of forming an efficient parliamentary administration to support the work of elected officials.[12]

Further in 2012, France and Afghanistan signed a friendship and cooperation treaty for 20 years establishing France's long-term commitment and demonstrating its shift from a predominantly military relationship to a predominantly civilian one.[13] It was supplemented by a five-year action plan which provides details on projects in the areas of security (military and police training), scientific, cultural, and technical cooperation (agriculture, research, education, health, archaeology, governance), infrastructure (irrigation, electrification), economics, and trade.[14]

**First step: Acknowledging the Taliban's representative**

Referencing Monica Duffy Toft, Max Fisher recently wrote in the *New York Times* that "victorious insurgents badly need 'international legitimacy, support and aid' in order to cement their rule."[15] While international legitimacy might be an aspiration for the Taliban's leadership, the more immediate concern is funding its nascent government. From non-Taliban bureaucrats such as the former American citizen and current mayor of Kabul, Daoud Sultanzoy, who has not been paid in months to villagers unable to buy food, the people and country of Afghanistan face a "severe economic crisis."[16][17] Underscoring the immediate need for foreign assistance or, at a minimum, access to the global financial infrastructure is the fact that "nearly 70 percent of the Afghan government's budget came from foreign aid."[18]

---

[11] Ibid.
[12] Ibid.
[13] *France and Afghanistan Bilateral relations* (accessed September 28, 2021); available from https://www.diplomatie.gouv.fr/en/country-files/afghanistan/.
[14] Ibid.
[15] Fisher, Max. "Seeking Global Recognition, Taliban Take a New Approach: Making Nice." *The New York Times*, August 19, 2021, sec. World. https://www.nytimes.com/2021/08/19/world/asia/taliban-foreign-recognition.html?action=click&module=RelatedLinks&pgtype=Article.

      The practical need to import food and maintain essential services has driven Taliban representatives to request international assistance and relief from U.S. sanctions that, among other restrictions, block the country's access to its foreign reserves held in the United States. While the United Nations has pledged aid to non-government organizations, the conditions, logistics, and diplomacy of rendering any assistance remain undetermined. The details and mechanics of organizing and coordinating humanitarian assistance provide an immediate opportunity for France to take a leadership role in any efforts. The Taliban recognizes France's global standing and even "before the last military flight lifted off from Kabul's airport…made an appeal for diplomatic relations" to the permanent member of the U.N. Security Council.[19] Perhaps in an opportunistic bid to (re)assert strategic autonomy, President Macron "has made a slew of phone calls to the leaders of Afghanistan's neighbors and regional influencers, including Iran, Iraq, Russia, Kazakhstan, Uzbekistan, and Tajikistan."[20] France's outreach to possible supporters is a step in the right direction for France to lead from the front.

---

[16] Yaroslav, Trofimov. "In Kabul, a Former American Citizen Keeps Running the City under Taliban Watch." *Wall Street Journal*, September 22, 2021, sec. World. https://www.wsj.com/articles/in-kabul-a-former-american-citizen-keeps-running-the-city-under-taliban-watch-11632303000.
[17] Yaroslav, "In Kabul, a Former American Citizen Keeps Running the City under Taliban Watch."
[18] Curtis, Lisa. "How America Should Deal with the Taliban." Foreign Affairs, October 1, 2021. https://www.foreignaffairs.com/articles/afghanistan/2021-09-20/how-america-should-deal-taliban.
[19] Tarquinio, J. Alex. "Macron Uses Biden's Afghan Retreat to Push 'Strategic Autonomy." Foreign Policy, September 2, 2021. https://foreignpolicy.com/2021/09/02/macron-biden-afghanistan-retreat-strategic-autonomy/.
[20] Tarquinio, "Macron Uses Biden's Afghan Retreat to Push 'Strategic Autonomy.'"

However, the republic must also balance its stated commitment to making humanitarian action a pillar of its foreign policy while not enabling violations of the basic rules of international humanitarian law.[21] These objectives are not mutually exclusive, and France must balance its domestic and international political priorities of self-assertion with humanitarian leadership. To start, France must clearly and forcefully communicate that formal diplomatic recognition, if ever feasible, is a mid-to-long-term process. Instead, as an initial pragmatic step of facilitating and coordinating immediate humanitarian assistance, France should conditionally acknowledge the Taliban's representation of the country's population of almost 40 million, 42% under the age of fourteen.[22] Top of mind should be the Taliban's not so distant history of reneging on cooperative efforts and quickly backtracking on "conciliatory remarks," which the organization used to ease international concerns about its criminal code and its harsh interpretation of Islam.[23] Though written before the withdrawal of U.S. forces, Barnett R. Rubin of the United States Institution of Peace notes that "the Taliban's quest for recognition and eventual eligibility for aid provides some of the most important leverage that other actors have over them.."[24] He continues that "bargaining over conditions for sanctions relief, recognition and assistance enjoys an important advantage over relying on military pressure."[25] France (and its partners) should tactfully employ this leverage to set non-negotiable, observable conditions tied to humanitarian aid and assistance and the framework for longer-term, more complex diplomatic negotiations. France must frame and establish these conditions in practical, focused parameters observable by neutral non-government agencies or regional democracies such as India.[26] Adopting language from the U.S. North Korean Human Rights Act of 2004, France and any partners should ensure that: assistance is delivered and monitored under international humanitarian standards, is provided on a need basis, directly reaches the intended beneficiaries, and that access is allowed to vulnerable groups, specifically women, minorities, and other populations, regardless of location.[27] The security, passage, and free external communications of any monitors or distributors of aid must be unconditionally guaranteed.

Further, there must be no discriminatory conditions (e.g., sexual orientation, gender, religion, or nationality) placed on assistance personnel or organizations. Finally, the assistance and direct distribution of medical care and supplies, food, and other aid will have a pre-determined duration. These conditions must be acknowledged, without reservation, within a finite window of time. The focus of assistance will be saving lives and supporting the Afghan population. Delay in accepting these requirements, posturing, or backchanneling by the Taliban must immediately be met with public condemnation, a cease of negotiations, and a path to increased international pressure. If the Taliban is sincere in helping the people of Afghanistan, then acceptance of France and the international community's offer of assistance cannot be quid pro quo.

**The Long Road to Recognition**

As noted earlier, the Taliban's desire for recognition from the international community provides France (and other foreign actors) an opportunity to take the lead in entering into practical cooperation for much needed humanitarian aid and potentially establishing diplomatic relations and recognizing it as a legitimate government. However, France's decision to recognize the Taliban government is influenced by domestic interests, particularly terrorism and immigration. These two issues are at the center of the debate leading up to the 2022 French Presidential elections, where Macron will again face far-right leader Marine Le Pen.

---

[26] Curtis, "How America Should Deal with the Taliban."
[27] U.S. Congress. *North Korean Human Rights Act of 2004.* Washington, DC: GPO:2004. https://www.congress.gov/108/crpt/hrpt478/CRPT-108hrpt478.pdf

Terrorism and Islamist radicalism are thorny issues in France. In the *Terrorism in Europe in 2020* report published by Europol, "E.U. Member States assessed that jihadist terrorism remained the greatest terrorist threat in the E.U."[28] From 2018 – 2020, France had the most jihadist attacks among E.U. member states and the U.K.[29] In addition, the fall of Kabul to the Taliban raises uneasy questions about its long-standing relationship with Al Qaeda. The Taliban sheltered Osama bin Laden before and after the 9/11 attacks, resulting in the US-led Afghanistan invasion in 2001.[30] While the Taliban has since expressed that Afghanistan will no longer become a haven for international terrorists, it appears the group has not broken ties with Al Qaeda.

Compounding the issue of terrorism is the presence of Islamic State Khorasan or ISIS-K in Afghanistan. The group was founded around six years ago by disaffected members of the Pakistani and Afghan Taliban.[31] A sworn enemy of the Taliban, ISIS-K competed with the Taliban over resources and ideological differences.[32] Over the years, the group has launched a series of attacks in Afghanistan, targeting the Taliban and civilians and security forces, including NATO

---

[28] Europol, "*European Union Terrorism Situation and Trend Report 2021*" (accessed September 29, 2021); available from https://www.europol.europa.eu/activities-services/main-reports/european-union-terrorism-situation-and-trend-report-2021-tesat, 13.
[29] Ibid., 16.
[30] Steven Erlanger, "Will the World Formally Recognize the Taliban?," *The New York Times*, September 1, 2021, https://www.nytimes.com/2021/09/01/world/asia/taliban-un-afghanistan-us.html.
[31] Joe Hernandez, "What We Know About ISIS-K, The Group Behind The Kabul Attack," *NPR*, August 28, 2021, https://www.npr.org/2021/08/26/1031349674/isis-k-taliban-who-what-you-need-to-know.
[32] Ibid.

troops.³³ Thus, Afghanistan once again becoming a haven for international terrorists threatens France's national security. Macron's foreign policy vis-à-vis Afghanistan is constrained by the upcoming election and the palpable fear of terrorism and Islamist radicalism among the French population, especially when the country is witnessing the trial for the November 2015 ISIS attacks in Paris. A threat assessment conducted by the Center for Strategic & International Studies points to ISIS-K stepping up attacks and recruitments efforts, the success of which will depend on "the Taliban's speed and success in establishing a government, local and regional counterterrorism efforts."³⁴ This assessment highlights how pressing the issue is for France and other foreign actors, given the Taliban's lack of counterterrorism capabilities.

Another contentious domestic issue for France is immigration. A functioning Afghan government that meets its population's basic needs will help mitigate a massive inflow of migrants and refugees into France and the rest of Europe. Migration is also a politically divisive topic in the country. President Emmanuel Macron faces Marine Le Pen once again in next year's presidential elections. Le Pen has been using her anti-immigration rhetoric to gain support from French voters. While Macron embraced stricter immigration measures in 2019 to regain supporters, an inflow of migrants and refugees from Afghanistan could potentially erode voter confidence.

While France's decision to hold dialogues with Taliban leaders aims to avoid a humanitarian crisis in Afghanistan in the immediate term, a lack of a functioning government in Afghanistan could still result in a refugee crisis in the years to come. While it can be argued that the withdrawal of the U.S. from Afghanistan accelerated the fall of the Afghan government, Europe will have to deal with the majority of Afghans seeking refuge from Taliban rule. France, along with Germany, has spoken up about finding a common response amidst growing concerns across the E.U. that there might be a repeat of the 2015 refugee crisis.[35]

---

³³ Catrina Doxsee, Jared Thompson, and Grace Hwang, "Examining Extremism: Islamic State Khorasan Province (ISKP)," Center for Strategic & Intternational Studies, September 8, 2021, https://www.csis.org/blogs/examining-extremism/examining-extremism-islamic-state-khorasan-province-iskp.
³⁴ Ibid.

**Conclusion**

One of the widely debated topics following the fall of the Afghan government to the Taliban on August 15, 2021, was whether the international community, particularly the U.S. and its European allies, would recognize the new regime. International recognition is something that the Taliban has always sought. Therefore, this paper sought to examine France's position vis-à-vis the recognition of the Taliban as a legitimate government. France maintains interests in Afghanistan and has been heavily involved militarily and in the country's reconstruction since 2002. Immediately following the fall of Kabul to the Taliban, France has been one of the prominent foreign actors that took the lead in initiating dialogues with the Taliban's representative. Through this practical cooperation, France could take a leadership role in organizing and coordinating humanitarian assistance to Afghanistan's population.

In the meantime, the Taliban takeover poses threats to France's domestic interests, particularly in the areas of immigration and terrorism. Recognizing the Taliban government might open diplomatic relations that allow the two countries to manage these issues better. As discussed elsewhere in this paper, France has a history of unilaterally recognizing the Afghanistan government following the Soviet invasion. That being said, we do not believe that France is close to recognizing the Taliban. Circumstances in Afghanistan are rapidly evolving, and France, as with the rest of the E.U., will continue its wait-and-see approach. For Macron, deciding on Taliban recognition at this point could potentially undermine his bid for the presidency. He must balance

---

[35] Silvia Amaro, "Europe fears a repeat of 2015 refugee crisis as Afghanistan collapses," *CNBC*, August 18, 2021, https://www.cnbc.com/2021/08/18/europe-fears-a-repeat-of-2015-refugee-crisis-as-afghanistan-collapses.html.

France's geopolitical interests against working with a militant government known for its human rights abuses and ties with international terrorists. While we recognize that France's decision to recognize the Taliban or not will also be influenced by changing global political dynamics, we believe that its near-term policy will be more influenced by domestic factors, as discussed above.